\documentclass[aps,prd,twocolumn,superscriptaddress,showpacs]{revtex4}
\usepackage{epsfig,epsf}
\usepackage{amsmath}
\usepackage{amsthm}
\usepackage{amsfonts}
\usepackage{amssymb}
\usepackage{dsfont}
\usepackage{multirow}
\usepackage{appendix}
\usepackage{slashed}
\usepackage[active]{srcltx}
\usepackage{psfrag}

\newcommand{\be}{\begin{equation}}
\newcommand{\ee}{\end{equation}}
\newcommand{\ba}{\begin{eqnarray}}
\newcommand{\ea}{\end{eqnarray}}
\newcommand{\baa}{\begin{eqnarray*}}
\newcommand{\btab}{\begin{tabular}}
\newcommand{\etab}{\end{tabular}}
\newcommand{\eaa}{\end{eqnarray*}}

\def\inbar{\,\vrule height1.5ex width.4pt depth0pt}
\def\IC{\relax\hbox{$\inbar\kern-.3em{\rm C}$}}
\def\IZ{\relax{\hbox{\cmss Z\kern-.4em Z}}}
\def\IR{{\hbox{{\rm I}\kern-.2em\hbox{\rm R}}}}

\def\IP{{\hbox{{\rm I}\kern-.2em\hbox{\rm P}}}}
\def\II{\hbox{{1}\kern-.25em\hbox{l}}}

\begin{document}

\title{ Strong $Z_c^{+}(3900)\rightarrow J/\psi \pi^{+}; \eta_{c} \rho^{+}$
decays in QCD}
\date{\today}
\author{S.~S.~Agaev}
\affiliation{Department of Physics, Kocaeli University, 41380 Izmit, Turkey}
\affiliation{Institute for Physical Problems, Baku State University, Az--1148 Baku,
Azerbaijan}
\author{K.~Azizi}
\affiliation{Department of Physics, Do\v{g}u\c{s} University, Acibadem-Kadik\"{o}y, 34722
Istanbul, Turkey}
\author{H.~Sundu}
\affiliation{Department of Physics, Kocaeli University, 41380 Izmit, Turkey}

\begin{abstract}
The widths of the strong decays $Z_c^{+}(3900)\to J/\psi \pi^{+}$ and $%
Z_c^{+}(3900)\to \eta_{c} \rho^{+}$ are calculated. To this end, the mass
and decay constant of the exotic $Z_c^{+}(3900)$ state are computed by means
of a two-point sum rule. The obtained results are then used to calculate the
strong couplings $g_{Z_{c}J/\psi \pi }$ and $g_{Z_{c}\eta_{c} \rho }$
employing QCD sum rules on the light-cone supplied by a technique of the
soft-meson approximation. We compare our predictions on the mass and decay
widths with available experimental data and other theoretical results.
\end{abstract}

\maketitle

\section{Introduction}

The exotic hadronic states, i.e. ones that can not be included into the
quark-antiquark and three-quark bound schemes of the standard spectroscopy
already attracted interests of physicists \cite%
{GellMann:1964nj,Witten:1979kh}.
The quantitative investigations of such states are connected with the
invention of the QCD sum rule method \cite{Shifman:1979}, which was employed
for analysis of glueballs, hybrid $q\overline{q}g$ resonances \cite%
{Shifman:1979,Balitsky:1982ps,Reinders:1985}, exotic four-quark $\bar {u}%
\bar {u}sd$ mesons \cite{Braun:1985ah,Braun:1988kv} and six-quark systems
\cite{Larin:1986}. But because of problems of old experiments, stemmed
mainly from difficulties in detecting heavy resonances, the existence of the
exotic states was not then certainly established.

The situation changed dramatically during the last decade when the Belle,
BaBar, LHCb and BES collaborations began copiously to yield experimental
data providing an information on the masses, decay width and quantum numbers
$J^{PC}$ of new exotic states. Starting from the discovery of the
charmonium-like resonance $X(3872)$ by Belle \cite{Belle:2003}, confirmed
later by other experiments \cite{D0:2004}, studying of new $XYZ$ family of
mesons became one of the interesting and rapidly growing branches of the
high energy physics (see, the reviews \cite%
{Swanson:2006st,Klempt:2007cp,Godfrey:2008nc,Voloshin:2007dx,Nielsen:2010,
Faccini:2012pj,Esposito:2014rxa} and references therein).

There were attempts to describe the new charmonium-like resonances as
excitations of the ordinary $c\overline{c}$ charmonium: In order to compute
the charmonium spectrum, various quark-antiquark potentials were used and
their mass and radiative transitions to other charmonium states were studied
\cite{Barnes:2005pb}. It should be noted that some of new resonances allow
interpretation as the excited $c\overline{c}$ states. But the bulk of the
collected experimental data can not be included into this scheme, and hence
for their explanation new-unconventional quark configurations are required.
To this end various quark-gluon models were suggested. They differ from each
other in elements of the substructure, and in mechanisms of the strong
interactions between these elements that form bound states.

One of the most employed in this context models is the four-quark or the
tetraquark model of the new resonances, already used to analyze the
light-quark exotic mesons \cite{Braun:1985ah,Braun:1988kv}. In the renewed
tetraquark model, the hadronic bound state is formed by two heavy and two
light quarks $Q\overline{Q}q\overline{q}$. These quarks may group into
compact tetraquark state, where all quarks have overlapping wave functions
\cite{Braaten:2013oba}. Their strong interaction can be studied by the quark
potential models that include not only the 2-body potentials of pairwise
quark interactions, but also the 3-body and 4-body potentials. In other
model four quarks cluster into the colored diquark $Qq$ and antidiquark $%
\overline {Q} \overline {q}$, which emerge as the elements of the
substructure. Diquark-antidiquarks are organized in such a way that
reproduce quantum numbers of the corresponding exotic states \cite%
{Maiani:2004vq}. In this picture the bound state forms due to not only the
quark-antiquark potentials, but also owing to the diquark-antidiquark
interactions. Alternatively, in the meson molecule picture, the quarks
appear as color-singlet $\overline{Q}q$ and $Q\overline{q}$ mesons. Finally,
in the hadro-quarkonium model suggested in Ref.\ \cite{Dubynskiy:2008mq},
the four-quarks create the bound system consisting of colorless $Q\overline{Q%
}$ and $q\overline{q}$ pairs of the heavy and light quarks, respectively. In
the molecule and hadro-quarkonium models the strong interactions are
mediated by the meson exchange.

Another possibility to describe the four-quark state is the Born-Oppenheimer
tetraquark structure proposed recently in Ref.\ \cite{Braaten:2013boa}. In
this approach the heavy quarks $Q$ and $\overline{Q}$ are considered as
being embedded in the configuration of gluon and light-quark fields, which
are not a flavor singlet, but have isospin 1. The exotic mesons can also be
considered in the framework of the traditional hybrid models, as particles
consisting of the heavy quarks and a gluon $Q\overline{Q}g$. The gluonic
excitation in this approach is treated as a constituent-particle with
definite quantum numbers. It should be noted, however that none of these
models firmly succeeded in analysis of the variety of the available
experimental data: in order to describe features of the observed exotic
states one should involve different models.

The $XYZ$ meson masses and decay widths were calculated, and their quantum
numbers $J^{PC}$ analyzed using all of the aforementioned models and various
theoretical methods. Theoretical approaches within QCD include the lattice
simulations to explore the exotic and excited charmonium spectroscopy \cite%
{Dudek:2007wv,Liu:2012ze,Moir:2013ub}, the calculations based on the
different quark potential models \cite{Swanson:2006st,Voloshin:2007dx}, and
QCD sum rule method (see, for review Ref.\ \cite%
{Voloshin:2007dx,Nielsen:2010}). The evaluation of the masses, decay
constants and widths of numerous exotic states and comparison of the
obtained results with the accumulated experimental data yield valuable
information on the quark-gluon structure of new states and mechanisms of the
strong interactions between their building elements. Despite remaining
problems, one can state that now important parts of the whole picture of
exotic multi-quark states are clearer than in the beginning of the decade.

The $Z_{c}^{\pm }$ states discovered by BESIII in the process $%
e^{+}e^{-}\rightarrow \pi ^{+}\pi ^{-}J/\psi$ \cite{Ablikim:2013mio}, were
observed by the Belle collaboration \cite{Liu:2013dau}, as well. Their
existence were also confirmed in Ref.\ \cite{Xiao:2013iha} on the basis of
the CLEO-c data analysis. The $Z_{c}^{\pm }\rightarrow J/\psi \pi ^{\pm }$
decays demonstrate that $Z_{c}^{\pm }$ are tetraquark states with
constituents $c\bar{c}u\bar{d}$ and $c\bar{c}d\bar{u}$.  Observation of the neutral
partner of $Z_{c}^{\pm }$ in the process
$e^{+}e^{-}\rightarrow \pi ^{0}Z_{c}^{0}\rightarrow \pi ^{0}\pi^{0}J/\psi $ was reported
in Ref.\ \cite{Ablikim:2015tbp}. Theoretical investigations of the $Z_c$ states
encompass different models and approaches (see, Refs.\
\cite{Wang:2010rt,Dias:2013xfa,Wang:2013vex,Deng:2014gqa,Esposito} and references therein).

In this work we evaluate the widths of the strong decays $Z_c^{+}(3900)\to J/\psi \pi^{+}$ and $%
Z_c^{+}(3900)\to \eta_{c} \rho^{+}$ of  $Z_{c}^{+}(3900)$ (in what follows denoted
as $Z_{c}$) considering it as the diquark-antidiquark state. For these purposes, we compute
the mass and decay constant of $Z_c$,
as well as the couplings of the strong $Z_{c}J/\psi \pi$ and $Z_{c}\eta_{c}\rho$
vertices allowing us to find the required decay widths. For calculation of the mass and
decay constant we employ QCD two-point sum rule, whereas in the
case of the strong couplings apply methods of QCD light-cone sum
rule (LCSR) supplemented by the soft-meson approximation
\cite{Braun:1989,Ioffe:1983ju,Braun:1995}. The latter is necessary because $Z_{c}$ state
contains the four valence quarks, as a result, the light-cone expansion of
the correlation functions inevitably reduces to the short-distance expansion
in terms of local matrix elements. In the context of LCSR approach this correspondences
to the vanishing meson momentum. In the present work we adopt the zero-momentum limit
for the mesons referring to the approach itself as the soft-meson approximation.
This approximation is rather simple, and, as we shall see, leads to nice agreement with
the experimental data \cite{Ablikim:2013mio,Liu:2013dau}. Within the sum rule method
$Z_c$ state was studied previously in Refs.\ \cite{Wang:2010rt,Dias:2013xfa,Wang:2013vex}.
Thus, in order to calculate strong couplings and decay widths of $Z_c$ state
in Ref.\ \cite{Dias:2013xfa} QCD three-point sum rule method was employed.

This article is organized in the following way. In section \ref{sec:Mass},
we calculate the mass and decay constant of the $Z_{c}$ state within
two-point QCD sum rule approach. Section \ref{sec:Vertex} is devoted to
calculation of the strong $Z_{c}J/\psi \pi$ and $Z_{c}\eta _{c}\rho $
vertices, where the sum rules for the couplings $g_{Z_{c}J/\psi \pi }$ and $%
g_{Z_{c}\eta _{c}\rho }$ are derived. Here we also calculate the widths of
the decay channels under consideration. Numerical computations of the mass,
decay constant, strong couplings, and decay widths are performed in Section %
\ref{sec:Num}. The obtained results are compared with the available
experimental data, as well as with existing theoretical calculations. This
Section contains also our conclusions. The explicit expression of the
spectral density $\rho ^{QCD}(s)$ necessary for computation of the mass and
decay constant of $Z_c$ state is moved to Appendix A \ref{sec:App}.


\section{The mass and decay constant of the $Z_c$ state}

\label{sec:Mass}

In order to calculate the mass and decay constant of the $Z_{c}^{+}$ state
in the framework of QCD sum rules, we start from the two-point correlation
function
\begin{equation}
\Pi _{\mu \nu }(q)=i\int d^{4}xe^{iq\cdot x}\langle 0|\mathcal{T}\{J_{\mu
}^{Z_c}(x)J_{\nu }^{Z_c\dagger }(0)\}|0\rangle ,  \label{eq:CorrF1}
\end{equation}%
where the interpolating current with required quantum numbers $J^{PC}=1^{+-}$
is given by the following expression
\begin{eqnarray}
J_{\nu }^{Z_{c}}(x) &=&\frac{i\epsilon \tilde{\epsilon}}{\sqrt{2}}\left\{ %
\left[ u_{a}^{T}(x)C\gamma _{5}c_{b}(x)\right] \left[ \overline{d}%
_{d}(x)\gamma _{\nu }C\overline{c}_{e}^{T}(x)\right] \right.  \notag \\
&&\left. -\left[ u_{a}^{T}(x)C\gamma _{\nu }c_{b}(x)\right] \left[ \overline{%
d}_{d}(x)\gamma _{5}C\overline{c}_{e}^{T}(x)\right] \right\} .
\label{eq:Curr}
\end{eqnarray}%
Here we have introduced the short-hand notations $\epsilon =\epsilon _{abc}$
and $\tilde{\epsilon}=\epsilon _{dec}$. In Eq.\ (\ref{eq:Curr}) $a,b,c,d,e$
are color indexes and $C$ is the charge conjugation matrix.

In order to derive QCD sum rule expression we first calculate the
correlation function in terms of the physical degrees of freedom. Performing
integral over $x$ in Eq.\ (\ref{eq:CorrF1}), we get
\begin{equation*}
\Pi _{\mu \nu }^{\mathrm{Phys}}(q)=\frac{\langle 0|J_{\mu
}^{Z_{c}}|Z_{c}(q)\rangle \langle Z_{c}(q)|J_{\nu }^{Z_{c}\dagger }|0\rangle
}{m_{Z_{c}}^{2}-q^{2}}+...
\end{equation*}%
where $m_{Z_{c}}$ is the mass of the $Z_{c}$ state, and dots stand for
contributions of the higher resonances and continuum states. We define the
decay constant $f_{Z_{c}}$ through the matrix element
\begin{equation}
\langle 0|J_{\mu }^{Z_{c}}|Z_{c}(q)\rangle =f_{Z_{c}}m_{Z_{c}}\varepsilon
_{\mu },  \label{eq:Res}
\end{equation}%
with $\varepsilon _{\mu }$ being the polarization vector of $Z_{c}$ state.
Then in terms of $m_{Z_{c}}$ and $f_{Z_{c}}$, the correlation function can
be written in the following form
\begin{equation}
\Pi _{\mu \nu }^{\mathrm{Phys}}(q)=\frac{m_{Z_{c}}^{2}f_{Z_{c}}^{2}}{%
m_{Z_{c}}^{2}-q^{2}}\left( -g_{\mu \nu }+\frac{q_{\mu }q_{\nu }}{%
m_{Z_{c}}^{2}}\right) +\ldots  \label{eq:CorM}
\end{equation}%
The Borel transformation applied to Eq.\ (\ref{eq:CorM}) yields%
\begin{equation}
\mathcal{B}_{q^{2}}\Pi _{\mu \nu }^{\mathrm{Phys}%
}(q)=m_{Z_{c}}^{2}f_{Z_{c}}^{2}e^{-m_{Z_{c}}^{2}/M^{2}}\left( -g_{\mu \nu }+%
\frac{q_{\mu }q_{\nu }}{m_{Z_{c}}^{2}}\right) +\ldots
\end{equation}

The same function in QCD side, $\Pi _{\mu \nu }^{\mathrm{QCD}}(q)$, has to
be determined employing of the quark-gluon degrees of freedom. To this end,
we contract the heavy and light quark fields and find
\begin{eqnarray}
&&\Pi _{\mu \nu }^{\mathrm{QCD}}(q)=-\frac{i}{2}\int d^{4}xe^{iqx}\epsilon
\tilde{\epsilon}\epsilon ^{\prime }\tilde{\epsilon}^{\prime }\left\{ \mathrm{%
Tr}\left[ \gamma _{5}\widetilde{S}_{u}^{aa^{\prime }}(x)\right. \right.
\notag \\
&&\left. \times \gamma _{5}S_{c}^{bb^{\prime }}(x)\right] \mathrm{Tr}\left[
\gamma _{\mu }\widetilde{S}_{c}^{e^{\prime }e}(-x)\gamma _{\nu
}S_{d}^{d^{\prime }d}(-x)\right]  \notag \\
&&-\mathrm{Tr}\left[ \gamma _{\mu }\widetilde{S}_{c}^{e^{\prime
}e}(-x)\gamma _{5}S_{d}^{d^{\prime }d}(-x)\right] \mathrm{Tr}\left[ \gamma
_{\nu }\widetilde{S}_{u}^{aa^{\prime }}(x)\right.  \notag \\
&&\times \left. \gamma _{5}S_{c}^{bb^{\prime }}(x)\right] -\mathrm{Tr}\left[
\gamma _{5}\widetilde{S}_{u}^{a^{\prime }a}(x)\gamma _{\mu }S_{c}^{b^{\prime
}b}(x)\right]  \notag \\
&&\times \mathrm{Tr}\left[ \gamma _{5}\widetilde{S}_{c}^{e^{\prime
}e}(-x)\gamma _{\nu }S_{d}^{d^{\prime }d}(-x)\right] +\mathrm{Tr}\left[
\gamma _{\nu }\widetilde{S}_{u}^{aa^{\prime }}(x)\right.  \notag \\
&&\left. \times \left. \gamma _{\mu }S_{c}^{bb^{\prime }}(x)\right] \mathrm{%
Tr}\left[ \gamma _{5}\widetilde{S}_{c}^{e^{\prime }e}(-x)\gamma
_{5}S_{d}^{d^{\prime }d}(-x)\right] \right\} ,  \label{eq:CorrF2}
\end{eqnarray}%
where%
\begin{equation*}
\widetilde{S}_{c(q)}^{ij}(x)=CS_{c(q)}^{ij\mathrm{T}}(x)C.
\end{equation*}%
Here the heavy-quark propagator $S_{c}^{ij}(x)$ is given by the expression
\cite{Reinders:1984sr}
\begin{eqnarray}
&&S_{c}^{ij}(x)=i\int \frac{d^{4}k}{(2\pi )^{4}}e^{-ikx}\left[ \frac{\delta
_{ij}\left( {\slashed k}+m_{c}\right) }{k^{2}-m_{c}^{2}}\right.  \notag \\
&&-\frac{gG_{ij}^{\alpha \beta }}{4}\frac{\sigma _{\alpha \beta }\left( {%
\slashed k}+m_{c}\right) +\left( {\slashed k}+m_{c}\right) \sigma _{\alpha
\beta }}{(k^{2}-m_{c}^{2})^{2}}  \notag \\
&&\left. +\frac{g^{2}}{12}G_{\alpha \beta }^{A}G^{A\alpha \beta }\delta
_{ij}m_{c}\frac{k^{2}+m_{c}{\slashed k}}{(k^{2}-m_{c}^{2})^{4}}+\ldots %
\right] .  \label{eq:Qprop}
\end{eqnarray}%
In Eq.\ (\ref{eq:Qprop}) the short-hand notation
\begin{equation*}
G_{ij}^{\alpha \beta }\equiv G_{A}^{\alpha \beta
}t_{ij}^{A},\,\,\,\,A=1,\,2\,\ldots 8,
\end{equation*}%
is used, where $i,\,j$ are color indexes, and $t^{A}=\lambda ^{A}/2$ with $%
\lambda ^{A}$ being the standard Gell-Mann matrices. The first term in Eq.\ (%
\ref{eq:Qprop}) is the free (perturbative) massive quark propagator, next
ones are nonperturbative gluon corrections. In the nonperturbative terms the
gluon field strength tensor $G_{\alpha \beta }^{A}\equiv G_{\alpha \beta
}^{A}(0)$ is fixed at $x=0.$

The light-quark propagator employed in our work reads
\begin{eqnarray}
&&S_{q}^{ij}(x)=i\frac{{\slashed x}}{2\pi ^{2}x^{4}}\delta _{ij}-\frac{m_{q}%
}{4\pi ^{2}x^{2}}\delta _{ij}-\frac{\langle q\overline{q}\rangle }{12}
\notag \\
&&\times \left( 1-i\frac{m_{q}}{4}{\slashed x}\right) \delta _{ij}-\frac{%
x^{2}}{192}m_{0}^{2}\langle q\overline{q}\rangle \left( 1-i\frac{m_{q}}{6}{%
\slashed x}\right) \delta _{ij}  \notag \\
&&-i\frac{gG_{ij}^{\alpha \beta }}{32\pi ^{2}x^{2}}\left[ \sigma _{\alpha
\beta }{\slashed x+\slashed x}\sigma _{\alpha \beta }\right] +\ldots .
\label{eq:qprop}
\end{eqnarray}%
The correlation function $\Pi _{\mu \nu }^{\mathrm{QCD}}(q)$ has also the
following decomposition over the Lorentz structures%
\begin{equation}
\Pi _{\mu \nu }^{\mathrm{QCD}}(q)=\Pi ^{\mathrm{QCD}}(q^{2})g_{\mu \nu }+%
\widetilde{\Pi }^{\mathrm{QCD}}(q^{2})q_{\mu }q_{\nu }.
\end{equation}%
The QCD sum rule expression for the mass and decay constant can be derived
after choosing the same structures in both $\Pi _{\mu \nu }^{\mathrm{Phys}%
}(q)$ and $\Pi _{\mu \nu }^{\mathrm{QCD}}(q)$. We choose to work with the
term $\sim q_{\mu }q_{\nu }$ and invariant function $\widetilde{\Pi }^{%
\mathrm{QCD}}(q^{2})$, which can be represented as the dispersion integral
\begin{equation}
\widetilde{\Pi }^{\mathrm{QCD}}(q^{2})=\int_{4m_{c}^{2}}^{\infty }\frac{\rho
^{\mathrm{QCD}}(s)}{s-q^{2}}+...,
\end{equation}%
where $\rho ^{\mathrm{QCD}}(s)$ is the corresponding spectral density.

The QCD sum rule calculations requires utilization of some consecutive
operations: we recall only the main steps in the computational scheme used
in the present work to derive the spectral density $\rho ^{\mathrm{QCD}}(s)$%
. Thus, having employed the transformation%
\begin{eqnarray}
&&\frac{1}{(x^{2})^{n}}=\int \frac{d^{D}t}{(2\pi )^{D}}e^{-it\cdot
x}i(-1)^{n+1}2^{D-2n}\pi ^{D/2}  \notag \\
&&\times \frac{\Gamma (D/2-n)}{\Gamma (n)}\left( -\frac{1}{t^{2}}\right)
^{D/2-n},
\end{eqnarray}%
we first replace, where it is necessary, $x_{\mu }$ by $-i\partial /\partial
q_{\mu }$, and calculate the $x$ integral. As a result, we get the delta
function with a combination of the momenta in its argument. This Dirac delta
is used to remove one of the momentum integrals. The remaining integrations
over $t$ and over the momentum  require invoking the Feynman parametrization and
performing rearrangements of denominators in obtained expressions. Then we
carry out integration over $t$ and perform the last integral over $k$ by means
of the formulas
\begin{equation}
\int d^{4}k\frac{1}{(k^{2}+L)^{\alpha }}=\frac{i\pi ^{2}(-1)^{\alpha }\Gamma
(\alpha -2)}{\Gamma (\alpha )[-L]^{\alpha -2}},  \label{eq:F1}
\end{equation}%
and
\begin{eqnarray}
&&\int d^{4}k\frac{k_{\mu }k_{\nu }}{\left(
k^{2}-2Akq+Aq^{2}-Bm_{c}^{2}\right) ^{\alpha }}  \notag \\
&=&\frac{i\pi ^{2}(-1)^{\alpha +1}\Gamma (\alpha -3)}{\Gamma (\alpha )\left[
-L\right] ^{\alpha -3}}\left[ \frac{g_{\mu \nu }}{2}+\frac{A^{2}(\alpha -3)}{%
L}q_{\mu }q_{\nu }\right].  \notag \\
&&  \label{eq:F2}
\end{eqnarray}%
In Eqs.\ (\ref{eq:F1}) and (\ref{eq:F2}) we use the notations
\begin{equation*}
A =2r(w+z-1),\,\,\, B=r(w+z)(w-1),
\end{equation*}%
and
\begin{eqnarray}
&&L =r^{2}(w-1) \left \{q^2wz(w+z-1)-m_c^{2}\frac{w+z}{r} \right\}, \notag \\
&&r=\frac{1}{w^{2}+(w+z)(z-1)}.
\end{eqnarray}
By applying the replacement
\begin{equation}
\Gamma \left( \frac{D}{2}-n\right) \left( -\frac{1}{L}\right) ^{\frac{D}{2}%
-n}\rightarrow \frac{(-1)^{n-1}}{(n-2)!}(-L)^{n-2}\ln (-L),
\end{equation}%
in the obtained expression, we get the imaginary part of the correlation
function. The remaining integrals over the Feynman parameters $w$ and $z$ in
some simple cases can be carried out explicitly, or kept in their original
form supplemented as a factor by the Heaviside function $\theta (L)$. The results of our
calculations of the spectral density $\rho ^{\mathrm{QCD}}(s)$ performed
within this scheme are collected in Appendix A.

Applying the Borel transformation on the variable $q^{2}$ to the invariant
amplitude $\widetilde{\Pi }^{\mathrm{QCD}}(q^{2})$ , equating the obtained
expression with the relevant part of $\mathcal{B}_{q^{2}}\Pi _{\mu \nu }^{%
\mathrm{Phys}}(q)$, and subtracting the continuum contribution, we finally
obtain the required sum rule. Thus, the mass of the $Z_{c}$ state can be
evaluated from the sum rule
\begin{equation}
m_{Z_{c}}^{2}=\frac{\int_{4m_{c}^{2}}^{s_{0}}dss\rho ^{\mathrm{QCD}%
}(s)e^{-s/M^{2}}}{\int_{4m_{c}^{2}}^{s_{0}}ds\rho (s)e^{-s/M^{2}}},
\end{equation}%
whereas to extract the numerical value of the decay constant $f_{Z_c}$ we
employ the formula
\begin{equation}
f_{Z_{c}}^{2}e^{-m_{Z_{c}}^{2}/M^{2}}=\int_{4m_{c}^{2}}^{s_{0}}ds\rho ^{%
\mathrm{QCD}}(s)e^{-s/M^{2}}.
\end{equation}%
The last two expressions are required sum rules to evaluate the $Z_{c}$
state's mass and decay constant, respectively.

\section{THE STRONG VERTICES $Z_{c}J/\protect\psi \protect\pi $ and $Z_{c}%
\protect\eta_{c} \protect\rho$}

\label{sec:Vertex}

This section is devoted to the calculation of the widths of the $%
Z_{c}\rightarrow J/\psi \pi$ and $Z_{c}\rightarrow \eta _{c}\rho$ decays. To
this end we calculate the strong couplings $g_{Z_{c}J/\psi \pi }$ and $%
g_{Z_{c}\eta_{c} \rho}$ using methods of the QCD sum rules on the light-cone
in conjunction with the soft-meson approximation.


\subsection{THE $Z_{c}J/\protect\psi \protect\pi $ VERTEX}

\begin{figure}
\centerline{
\begin{picture}(210,170)(0,0)
\put(-15,5){\epsfxsize8.2cm\epsffile{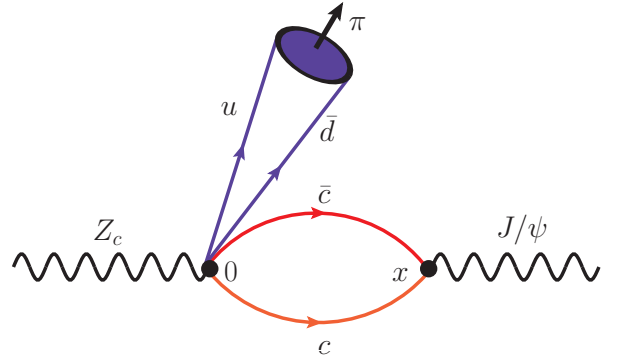}}
\end{picture}}
\caption{Sample diagram for the decay $Z_{c} \to J/\psi \pi$.}
\label{fig:SoftD}
\end{figure}

We start our analysis from the vertex $Z_{c}J/\psi \pi $ aiming to calculate
$g_{Z_{c}J/\psi \pi }$: we consider the correlation function
\begin{equation}
\Pi _{\mu \nu }(p,q)=i\int d^{4}xe^{ipx}\langle \pi (q)|\mathcal{T}\{J_{\mu
}^{J/\psi }(x)J_{\nu }^{Z_c\dagger }(0)\}|0\rangle ,  \label{eq:CorrF3}
\end{equation}%
where%
\begin{equation}
J_{\mu }^{J/\psi }(x)=\overline{c}_{i}(x)\gamma _{\mu }c_{i}(x),
\end{equation}%
and $J_{\nu }^{Z_{c}}(x)$ is defined by Eq.\ (\ref{eq:Curr}). Here $p$, $q$
and $p^{\prime }=p+q$ are the momenta of $J/\psi $, $\pi $ and $Z_{c}$,
respectively. A sample diagram describing the process $Z_{c} \to J/\psi \pi$
is depicted in Fig. \ref{fig:SoftD}.

To derive the sum rules for the coupling, we calculate $\Pi _{\mu \nu }(p,q)$
in terms of the physical degrees of freedom. Then it is not difficult to
obtain
\begin{eqnarray}
\Pi _{\mu \nu }^{\mathrm{Phys}}(p,q) &=&\frac{\langle 0| J_{\mu }^{J/\psi }|
J/\psi \left( p\right) \rangle }{p^{2}-m_{J/\psi }^{2}}\langle J/\psi \left(
p\right) \pi (q)| Z_{c}(p^{\prime })\rangle  \notag \\
&&\times \frac{\langle Z_{c}(p^{\prime })| J_{\nu }^{Z_c\dagger}| 0\rangle }{%
p^{\prime 2}-m_{Z_{c}}^{2}}+ \ldots .  \label{eq:CorrF4}
\end{eqnarray}%
where the dots denote contribution of the higher resonances and continuum
states.

We introduce the matrix elements
\begin{eqnarray}
&&\langle 0|J_{\mu }^{J/\psi }|J/\psi \left( p\right) \rangle =f_{J/\psi
}m_{J/\psi }\varepsilon _{\mu },  \notag \\
&&\langle Z_{c}(p^{\prime })|J_{\nu }^{Z_{c}\dagger}|0\rangle
=f_{Z_{c}}m_{Z_{c}}\varepsilon _{\nu }^{\prime \ast }  \notag \\
&&\langle J/\psi \left( p\right) \pi (q)|Z_{c}(p^{\prime })\rangle =\left[
(p\cdot p^{\prime })(\varepsilon ^{\ast }\cdot \varepsilon ^{\prime })\right.
\notag \\
&&\left. -(p\cdot \varepsilon ^{\prime })(p^{\prime }\cdot \varepsilon
^{\ast })\right] g_{Z_{c}J/\psi \pi },  \label{eq:Mel}
\end{eqnarray}%
where $f_{J/\psi },\,m_{J/\psi },\,\varepsilon _{\mu }$ are the decay
constant, mass and polarization vector of the $J/\psi $ meson, and $%
\varepsilon _{\nu }^{\prime }$ is the polarization vector of the $Z_{c}$
state. It is worth noting that the matrix element in the last row of Eq.\ (%
\ref{eq:Mel}) is defined in the gauge-invariant form.

Having used these matrix elements we can rewrite the correlation function as
\begin{eqnarray}
&&\Pi _{\mu \nu }^{\mathrm{Phys}}(p,q)=\frac{f_{J/\psi
}f_{Z_{c}}m_{Z_{c}}m_{J/\psi }g_{Z_{c}J/\psi \pi }}{\left( p^{\prime
2}-m_{Z_{c}}^{2}\right) \left( p^{2}-m_{J/\psi }^{2}\right) }  \notag \\
&&\times \left( \frac{m_{Z_{c}}^{2}+m_{J/\psi }^{2}}{2}g_{\mu \nu }-p_{\mu
}^{\prime }p_{\nu }\right) + \ldots  \notag \\
&&=\Pi _{\pi }^{\mathrm{Phys}}(p^{2},(p+q)^{2})g_{\mu \nu }+\widetilde{\Pi }%
_{\pi }^{\mathrm{Phys}}(p^{2},(p+q)^{2})p_{\mu }^{\prime }p_{\nu }.  \notag
\\
&& {}  \label{eq:CorrF5}
\end{eqnarray}%
For calculation of the strong coupling under consideration we choose to work
with the structure $\sim g_{\mu \nu }$. Then, for the corresponding
invariant function, we get
\begin{eqnarray}
&&\Pi _{\pi }^{\mathrm{Phys}}(p^{2},(p+q)^{2})=\frac{f_{J/\psi
}f_{Z_{c}}m_{Z_{c}}m_{J/\psi }g_{Z_{c}J/\psi \pi }}{\left( p^{\prime
2}-m_{Z_{c}}^{2}\right) \left( p^{2}-m_{J/\psi }^{2}\right) }  \notag \\
&&\times \frac{m_{Z_{c}}^{2}+m_{J/\psi }^{2}}{2}+\Pi ^{\mathrm{(RS:C)}%
}(p^{2},(p+q)^{2}),  \label{eq:CorrF5A}
\end{eqnarray}%
where $\Pi ^{\mathrm{(RS:C)}}(p^{2},(p+q)^{2})$ is the contribution arising
from the higher resonances and continuum states, that can be written down as
the double dispersion integral:
\begin{eqnarray}
&&\Pi ^{\mathrm{(RS:C)}}(p^{2},(p+q)^{2})=\int \int \frac{\rho
^{h}(s_{1},s_{2})ds_{1}ds_{2}}{(s_{1}-p^{2})(s_{2}-p^{\prime 2})}  \notag \\
&&+\int \frac{\rho _{1}^{h}(s_{1})ds_{1}}{(s_{1}-p^{2})}+\int \frac{\rho
_{2}^{h}(s_{2})ds_{2}}{(s_{2}-p^{\prime 2})}.
\end{eqnarray}%
This formula contains also single dispersion integrals that are necessary to
make the whole expression finite: As we shall see below, they play an
important role in the soft-meson approximation adopted in the present work.

In the standard LCSR approach in order to get sum rules for the strong
couplings \cite{Braun:1995,Aliev:1996,Khodjamirian:1999} one applies to Eq.\
(\ref{eq:CorrF5A}) double Borel transformation in variables $p^{2}$ and $%
p^{\prime 2}$ that vanishes the single dispersion integrals leaving in the
physical side of the LCSR only contributions of the ground state and the
double spectral integral. In other words, effects of higher resonances and
continuum states on the sum rule are under control and modeled by Borel
transformed double integral. Then, using the quark-hadron duality assumption
one replaces the spectral density $\rho^{h}(s_1,s_2)$ by its theoretical
counterpart $\rho^{QCD}(s_1,s_2)$, and subtracts the contribution of the
resonance and continuum states from the theoretical side of the sum rules.

But in the case under consideration, the situation differs from the standard
one. In fact, calculation of the function $\Pi _{\mu \nu }^{\mathrm{QCD}%
}(p,q)$ in the context of the perturbative QCD reveals its interesting
features. As is seen from Eqs.\ (\ref{eq:Curr}) and(\ref{eq:CorrF3}), the
tetraquark state contains four quarks at the same space-time location,
therefore contractions of the $c$ and $\overline{c}$ quark fields given at $%
x=0$ with the relevant fields at $x $ from the $J/\psi $ meson yield
expressions where the remaining light quarks are sandwiched between the pion
and vacuum states forming local matrix elements. In other words, we
encounter with the situation when dependence of the correlation function on
the meson distribution amplitudes disappears and integrals over the meson
DAs reduce to overall normalization factors. Within framework of LCSR method
such situation is possible in the kinematical limit $q \to 0$, when the light-cone
expansion reduces to the short-distant one. As a result, instead of the expansion in
terms of DAs one gets expansion over the local matrix elements \cite{Braun:1995}.
In this limit $p^{\prime}=p$ and relevant invariant amplitudes in the correlation
function depend only on one variable $p^2$. Here we adopt this approach, and following
Ref. \ \cite{Braun:1995} refer to the limit $q \to 0$ as the soft-meson approximation
bearing in mind that it actually implies calculation of the correlation function with
the equal initial and final momenta $p^{\prime}=p$, and dealing with the obtained double pole terms.

The soft-meson approximation considerably simplifies the QCD side of the sum rules,
but leads to more complicated expression for its hadronic representation. In the soft $%
p^{\prime }$\ $\rightarrow p$ limit, as it has been just emphasized above, the ground
state contribution depends only on the variable $p^{2}$. With some accuracy, it can be
written in the form
\begin{equation}
\Pi _{\pi }^{\mathrm{Phys}}(p^{2})=\frac{f_{J/\psi
}f_{Z_{c}}m_{Z_{c}}m_{J/\psi }g_{Z_{c}J/\psi \pi }}{\left(
p^{2}-m^{2}\right) ^{2}}m^{2},
\end{equation}%
where $m^{2}=\left( m_{Z_{c}}^{2}+m_{J/\psi }^{2}\right) /2$. The Borel
transformation in the variable $p^{2}$ applied to this correlation function
yields
\begin{eqnarray}
&&\Pi _{\pi }^{\mathrm{Phys}}(M^{2})=f_{J/\psi }f_{Z_{c}}m_{Z_{c}}m_{J/\psi
}g_{Z_{c}J/\psi \pi }  \notag \\
&&\times \frac{1}{M^{2}}e^{-m^{2}/M^{2}}m^{2}.
\end{eqnarray}%
But because within the soft-meson approximation we employ the one-variable
Borel transformation, now the single dispersion integrals also contribute to
the hadronic part of the sum rules. Non-vanishing contributions correspond
to transitions from the exited states in the $J/\psi $ channel with $m^{\ast
}>m_{J/\psi }$ to the ground state $J/\psi $ (the similar arguments are
valid for the $Z_{c}$ channel, as well). They are not suppressed relative to
the ground state contribution even after the Borel transformation \cite%
{Braun:1995,Ioffe:1983ju}. Hence, taking into account all unsuppressed
contributions to $\Pi ^{\mathrm{Phys}}(M^{2})$, denoted below as $A$, the
hadronic part of the sum rules can be schematically written in the form \cite%
{Braun:1995}
\begin{eqnarray}
&&\Pi _{\pi }^{\mathrm{Phys}}(M^{2})\simeq \frac{1}{M^{2}}\left\{ f_{J/\psi
}f_{Z_{c}}m_{Z_{c}}m_{J/\psi }m^{2}g_{Z_{c}J/\psi \pi }\right.  \notag \\
&&\left. +AM^{2}\right\} e^{-m^{2}/M^{2}}.
\end{eqnarray}%
It is evident, that the terms $\sim A$ emerge here as an undesired
contamination and make extraction of the strong coupling problematic. In
order to remove them from the hadronic part it is convenient to follow a
prescription suggested in Ref.\ \cite{Ioffe:1983ju} and act by the operator
\begin{equation}
\left( 1-M^{2}\frac{d}{dM^{2}}\right) M^{2}e^{m^{2}/M^{2}}  \label{eq:softop}
\end{equation}%
to both sides of the sum rules.

Now we need to calculate the correlation function in terms of the
quark-gluon degrees of freedom and find the QCD side of the sum rules.
Having contracted heavy quarks fields we get%
\begin{eqnarray}
&&\Pi _{\mu \nu }^{\mathrm{QCD}}(p,q)=\int d^{4}xe^{ipx}\frac{\epsilon
\widetilde{\epsilon }}{\sqrt{2}}\left[ \gamma _{5}\widetilde{S}%
_{c}^{ib}(x){}\gamma _{\mu }\right.  \notag \\
&&\left. \times \widetilde{S}_{c}^{ei}(-x){}\gamma _{\nu }+\gamma _{\nu }%
\widetilde{S}_{c}^{ib}(x){}\gamma _{\mu }\widetilde{S}_{c}^{ei}(-x){}\gamma
_{5}\right] _{\alpha \beta }  \notag \\
&&\times \langle \pi (q)|\overline{u}_{\alpha }^{a}(0)d_{\beta
}^{d}(0)|0\rangle ,  \label{eq:CorrF6}
\end{eqnarray}%
where $\alpha $ and $\beta $ are the spinor indexes.

To proceed we use the expansion
\begin{equation}
\overline{u}_{\alpha }^{a}d_{\beta }^{d}\rightarrow \frac{1}{4}\Gamma
_{\beta \alpha }^{j}\left( \overline{u}^{a}\Gamma ^{j}d^{d}\right) ,
\label{eq:MatEx}
\end{equation}%
where $\Gamma ^{j}$ is the full set of Dirac matrixes
\begin{equation*}
\Gamma ^{j}=\mathbf{1,\ }\gamma _{5},\ \gamma _{\lambda },\ i\gamma
_{5}\gamma _{\lambda },\ \sigma _{\lambda \rho }/\sqrt{2},
\end{equation*}%
Within the LCSR method we have also to use the light-cone expansion for the $%
c$-quark propagator
\begin{eqnarray}
&&\langle 0|\mathcal{T}\{c(x)\overline{c}(0)\}|0\rangle =i\int \frac{d^{4}k}{%
(2\pi )^{4}}e^{-ikx}\frac{{\slashed k}+m_{c}}{k^{2}-m_{c}^{2}}  \notag \\
&&-ig_{s}\int \frac{d^{4}k}{(2\pi )^{4}}e^{-ikx}\int_{0}^{1}dv\left[ \frac{1%
}{2}\frac{{\slashed k}+m_{c}}{\left( k^{2}-m_{c}^{2}\right) ^{2}}G^{\mu \nu
}\left( vx\right) \sigma _{\mu \nu }\right.  \notag \\
&&\left. -\frac{{\slashed k}+m_{c}}{k^{2}-m_{c}^{2}}vx_{\mu }G^{\mu \nu
}\left( vx\right) \gamma _{\nu }\right] +\ldots  \label{eq:Cprop}
\end{eqnarray}%
But because the light quark fields in the local matrix elements are fixed at
the point $x=0$, in the second piece of the propagator [Eq.\ (\ref{eq:Cprop}%
)] we expand the gluon field strength tensor at $x=0$ keeping only the
leading order term that is equivalent to usage the first two terms from $%
S_{c}^{ij}(x)$ (see, Eq.\ (\ref{eq:Qprop})) in the calculations .

In order to determine the required local matrix elements we consider first
the perturbative components of the heavy-quark propagators. To this end, it
is convenient to take sums over the color indexes. Using the overall color
factor $\epsilon _{abc}\epsilon _{dec}$, color factors of the propagators,
as well as the projector onto a color-singlet state $\delta ^{ad}/3$, it is
easy to demonstrate that for the perturbative contribution, the replacement
\begin{equation}
\frac{1}{4}\Gamma _{\beta \alpha }^{j}\left( \overline{u}^{a}\Gamma
^{j}d^{d}\right) \rightarrow \frac{1}{2}\Gamma _{\beta \alpha }^{j}\left(
\overline{u}\Gamma ^{j}d\right) .
\end{equation}%
is legitimate. For nonperturbative contributions, forming as a product of
the perturbative part of one of the propagators with the term $\sim G$ from
the other one, we find, for example,
\begin{equation*}
\epsilon _{abc}\epsilon _{dec}\delta _{ei}G_{ib}^{\rho \delta }\frac{1}{4}%
\Gamma _{\beta \alpha }^{j}\left( \overline{u}^{a}\Gamma ^{j}d^{d}\right)
\rightarrow -\frac{1}{4}\Gamma _{\beta \alpha }^{j}\left( \overline{u}\Gamma
^{j}G^{\rho \delta }d\right) .
\end{equation*}%
This rule allows us to insert into quark matrix elements the gluon field
strength tensor $G$ that effectively leads to three-particle components and
corresponding matrix elements of the pion. We neglect the terms $\sim GG$ appearing from the product of one-gluon components of the heavy propagators. Having finished a
color summation one can calculate the traces over spinor indexes.

The spectral density $\rho _{\pi }^{QCD}(s)$ has been found employing the
approach outlined in the Section \ref{sec:Mass}. Calculations demonstrate
that the pion local matrix element that, in the soft-meson limit,
contributes to the both structures of $\mathrm{Im}\Pi _{\mu \nu }^{\mathrm{%
QCD}}(p,q=0)$ is
\begin{equation}
\langle 0|\overline{d}(0)i\gamma _{5}u(0)|\pi (q)\rangle =f_{\pi }\mu _{\pi
},  \label{eq:MatE2}
\end{equation}%
where
\begin{equation}
\mu _{\pi }=\frac{m_{\pi }^{2}}{m_{u}+m_{d}}=-\frac{2\langle \overline{q}%
q\rangle }{f_{\pi }^{2}}.  \label{eq:PionEl}
\end{equation}%
The second equality in Eq.\ (\ref{eq:PionEl}) is the relation between $%
m_{\pi } $, $f_{\pi }$, the quark masses and the quark condensate $\langle
\overline{q}q\rangle $ arising from the partial conservation of axial
vector current (PCAC).

Choosing the structure $\sim g_{\mu \nu }$ for our analysis it is
straightforward to derive the corresponding spectral density
\begin{equation}
\rho _{\pi }^{\mathrm{QCD}}(s)=\frac{f_{\pi }\mu _{\pi }(s+2m_{c}^{2})\sqrt{%
s(s-4m_{c}^{2})}}{12\sqrt{2}\pi ^{2}s}.
\end{equation}
The continuum contribution can be subtracted in a standard manner after $%
\rho ^{h}(s)\rightarrow \rho _{\pi }^{\mathrm{QCD}}(s)$ replacement. The
final sum rule to evaluate the strong coupling reads
\begin{eqnarray}
&&g_{Z_{c}J/\psi \pi }=\frac{2}{f_{J/\psi }f_{Z_{c}}m_{Z_{c}}m_{J/\psi
}(m_{Z_{c}}^{2}+m_{J/\psi }^{2})}  \notag \\
&&\times \left( 1-M^{2}\frac{d}{dM^{2}}\right) M^{2}  \notag \\
&&\times \int_{4m_{c}^{2}}^{s_{0}}dse^{(m_{Z_{c}}^{2}+m_{J/\psi
}^{2}-2s)/2M^{2}}\rho _{\pi }^{\mathrm{QCD}}(s).  \label{eq:SRules}
\end{eqnarray}

The width of the decay $Z_{c}\rightarrow J/\psi \pi$ can be found applying
the standard methods and definitions for the strong coupling alongside with
other matrix elements [Eq.\ (\ref{eq:Mel})] and parameters of the $Z_{c}$
state. Our calculations give%
\begin{eqnarray}
&&\Gamma \left( Z_{c}\rightarrow J/\psi \pi\right) =\frac{g_{Z_{c}J/\psi \pi
}^{2}m_{J/\psi }^{2}}{24\pi }\lambda \left( m_{Z_{c}},\ m_{J/\psi },m_{\pi
}\right)  \notag \\
&&\times \left[ 3+\frac{2\lambda ^{2}\left( m_{Z_{c}},\ m_{J/\psi },m_{\pi
}\right) }{m_{J/\psi }^{2}}\right] ,  \label{eq:DW}
\end{eqnarray}%
where
\begin{equation*}
\lambda (a,\ b,\ c)=\frac{\sqrt{a^{4}+b^{4}+c^{4}-2\left(
a^{2}b^{2}+a^{2}c^{2}+b^{2}c^{2}\right) }}{2a}.
\end{equation*}%
The final expressions (\ref{eq:SRules}) and (\ref{eq:DW}) will be used for
numerical analysis of the decay channel $Z_{c}\rightarrow J/\psi \pi$.


\subsection{THE $Z_{c}\protect\eta _{c}\protect\rho $ VERTEX}


The coupling $g_{Z_{c}\eta _{c}\rho }$ can be calculated utilizing the
correlation function
\begin{equation}
\Pi _{\nu }(p,q)=i\int d^{4}xe^{ipx}\langle \rho (q)|\mathcal{T}\{J^{\eta
_{c}}(x)J_{\nu }^{Z_{c}\dagger }(0)\}|0\rangle ,
\end{equation}%
where the current $J^{\eta _{c}}(x)$ is defined as%
\begin{equation*}
J^{\eta _{c}}(x)=\overline{c}_{i}(x)i\gamma _{5}c_{i}(x).
\end{equation*}%
In order to find the hadronic representation of the correlation function we
define the matrix element:
\begin{equation}
\langle 0|J^{\eta _{c}}|\eta _{c}(p)\rangle =\frac{f_{\eta _{c}}m_{\eta
_{c}}^{2}}{2m_{c}},
\end{equation}%
with $m_{\eta _{c}}$ and $f_{\eta _{c}}$ being the $\eta _{c}$ meson's mass
and decay constant. The vertex $Z_{c}\eta _{c}\rho $ is defined as in Eq.\ (%
\ref{eq:Mel})
\begin{eqnarray}
&&\langle \eta _{c}\left( p\right) \rho (q)|Z_{c}(p^{\prime })\rangle =
\left[ (q\cdot p^{\prime })(\varepsilon ^{\ast }\cdot \varepsilon ^{\prime
})\right.  \notag \\
&&-\left. (q\cdot \varepsilon ^{\prime })(p^{\prime }\cdot \varepsilon
^{\ast })\right] g_{Z_{c}\eta _{c}\rho },
\end{eqnarray}%
but with $q$ and $\varepsilon $ being now the momentum and polarization
vector of the $\rho $-meson, respectively.

Then the calculation of the hadronic representation $\Pi _{\nu }^{\mathrm{%
Phys}}(p,q)$ is straightforward and yields
\begin{eqnarray}
\Pi _{\nu }^{\mathrm{Phys}}(p,q) &=&\frac{\langle 0|J^{\eta _{c}}|\eta
_{c}\left( p\right) \rangle }{p^{2}-m_{\eta _{c}}^{2}}\langle \eta
_{c}\left( p\right) \rho (q)|Z_{c}(p^{\prime })\rangle  \notag \\
&&\times \frac{\langle Z_{c}(p^{\prime })|J_{\nu }^{Z_{c}\dagger}|0\rangle }{%
p^{\prime 2}-m_{Z_{c}}^{2}}+\ldots .
\end{eqnarray}%
Employing the corresponding matrix elements we find for the ground state
contribution
\begin{eqnarray}
&&\Pi _{\nu }^{\mathrm{Phys}}(p,q)=\frac{f_{\eta
_{c}}f_{Z_{c}}m_{Z_{c}}m_{\eta _{c}}^{2}g_{Z_{c}\eta _{c}\rho }}{%
2m_{c}\left( p^{\prime 2}-m_{Z_{c}}^{2}\right) \left( p^{2}-m_{\eta
_{c}}^{2}\right) }  \notag \\
&&\times \left( \frac{m_{\eta _{c}}^{2}-m_{Z_{c}}^{2}}{2}\epsilon _{\nu
}^{\ast }+p^{\prime }\cdot \epsilon ^{\ast }q_{\nu }\right) +\ldots
\end{eqnarray}%
In the soft-meson limit only the structure $\sim \epsilon _{\nu }^{\ast }$
survives. The relevant invariant amplitude is given by the formula%
\begin{equation}
\Pi _{\rho }^{\mathrm{Phys}}(p^{2})=\frac{f_{\eta
_{c}}f_{Z_{c}}m_{Z_{c}}m_{\eta _{c}}^{2}g_{Z_{c}\eta _{c}\rho }}{%
4m_{c}\left( p^{2}-\widetilde{m}^{2}\right) ^{2}}(m_{\eta
_{c}}^{2}-m_{Z_{c}}^{2})+\ldots
\end{equation}%
where $\widetilde{m}^{2}=(m_{\eta _{c}}^{2}+m_{Z_{c}}^{2})/2$. The Borel
transformation of $\Pi _{\rho }^{\mathrm{Phys}}(p^{2})$ yields%
\begin{eqnarray}
&&\Pi _{\rho }^{\mathrm{Phys}}(M^{2})=\frac{f_{\eta
_{c}}f_{Z_{c}}m_{Z_{c}}m_{\eta _{c}}^{2}g_{Z_{c}\eta _{c}\rho }}{4m_{c}}%
(m_{\eta _{c}}^{2}-m_{Z_{c}}^{2})  \notag \\
&&\times \frac{1}{M^{2}}e^{-(m_{\eta _{c}}^{2}+m_{Z_{c}}^{2})/2M^{2}}+\ldots
\end{eqnarray}

Computation of the vertex $Z_{c}\eta _{c}\rho $ in terms of the quark-gluon
degrees of freedom is the next step to get the required sum rule. For the
correlation function $\Pi _{\nu }^{\mathrm{QCD}}(p,q)$ we obtain :
\begin{eqnarray}
&&\Pi _{\nu }^{\mathrm{QCD}}(p,q)=-i\int d^{4}xe^{ipx}\frac{\epsilon
\widetilde{\epsilon }}{\sqrt{2}}\left[ \gamma _{5}\widetilde{S}%
_{c}^{ib}(x){}\gamma _{5}\right.  \notag \\
&&\left. \times \widetilde{S}_{c}^{ei}(-x){}\gamma _{\nu }+\gamma _{\nu }%
\widetilde{S}_{c}^{ib}(x){}\gamma _{5}\widetilde{S}_{c}^{ei}(-x){}\gamma _{5}%
\right] _{\alpha \beta }  \notag \\
&&\times \langle \rho (q)|\overline{u}_{\alpha }^{d}(0)d_{\beta
}^{a}(0)|0\rangle .  \label{eq:CorrF7}
\end{eqnarray}%
In the soft-meson limit only the matrix elements
\begin{eqnarray}
&&\langle 0|\overline{u}(0)\gamma _{\mu }d(0)|\rho (p,\lambda )\rangle
=\epsilon _{\mu }^{(\lambda )}f_{\rho }m_{\rho },  \notag \\
&&\langle 0|\overline{u}(0)g\widetilde{G}_{\mu \nu }\gamma _{\nu }\gamma
_{5}d(0)|\rho (p,\lambda )\rangle =f_{\rho }m_{\rho }^{3}\epsilon _{\mu
}^{(\lambda )}\zeta _{4}  \label{eq:Mel2}
\end{eqnarray}%
contribute to $\mathrm{Im}\Pi _{\nu }^{\mathrm{QCD}}(p,q=0)$. The last
equality in Eq.\ (\ref{eq:Mel2}) is the matrix element of the twist-4
operator \cite{Ball:1998ff}: numerical value of the parameter $\zeta _{4}$
was evaluated within QCD sum rule at the renormalization scale $\mu =1\,\,{%
\mathrm{GeV}}$ in Ref.\ \cite{Braun:1985ah} and was found to be equal to $%
\zeta_{4}=0.15\pm 0.10$.

The spectral density $\rho _{\rho }^{\mathrm{QCD}}(s)$ can be derived in
accordance with the prescription described above. As a result we get:
\begin{equation}
\rho _{\rho }^{\mathrm{QCD}}(s)=\frac{f_{\rho }m_{\rho }\sqrt{s(s-4m_{c}^{2})%
}}{8\sqrt{2}\pi ^{2}}\left( 1+\frac{\zeta _{4}m_{\rho }^{2}}{s}\right) .
\end{equation}
Applying the operator from Eq.\ (\ref{eq:softop}) and equating the physical
representation for the invariant amplitude with its QCD expression we,
finally obtain the sum rule to evaluate the strong coupling $%
g_{Z_{c}\eta_{c}\rho}$. The width of the decay $Z_{c} \to \eta_{c} \rho$ is
given by Eq.\ (\ref{eq:DW}) with replacements $m_{\pi} \to m_{\eta_{c}}$ and
$m_{J/\psi} \to m_{\rho}$.

\section{Numerical computations and conclusions}

\label{sec:Num}
\begin{table}[tbp]
\begin{tabular}{|c|c|}
\hline\hline
Parameters & Values \\ \hline\hline
$m_{c}$ & $(1.275\pm0.025)~\mathrm{GeV}$ \\
$m_{J/\psi} $ & $(3096.92\pm0.01)~\mathrm{MeV}$ \\
$m_{\eta_{c}} $ & $(2983.6\pm 0.6) ~\mathrm{MeV} $ \\
$m_{\pi} $ & $(139.57018\pm 0.00035)~\mathrm{MeV} $ \\
$m_{\rho} $ & $(775.26\pm 0.25) ~\mathrm{MeV} $ \\
$f_{J/\psi} $ & $0.405 ~ \mathrm{GeV}$ \\
$f_{\eta_{c}} $ & $0.35 ~\mathrm{GeV}$ \\
$f_{\pi} $ & $131.5~\mathrm{MeV}$ \\
$f_{\rho} $ & $157$ $\mathrm{MeV}$ \\
$\langle \bar{q}q \rangle $ & $(-0.24\pm 0.01)^3 $ $\mathrm{GeV}^3$ \\
$\langle\frac{\alpha_sG^2}{\pi}\rangle $ & $(0.012\pm0.004)$ $~\mathrm{GeV}%
^4 $ \\
$m_0^2 $ & $(0.8\pm0.1) $ $\mathrm{GeV}^3$ \\ \hline\hline
\end{tabular}%
\caption{Input parameters used in calculations}
\label{tab:Param}
\end{table}

\begin{figure}
\centerline{
\begin{picture}(210,170)(0,0)
\put(-15,5){\epsfxsize8.2cm\epsffile{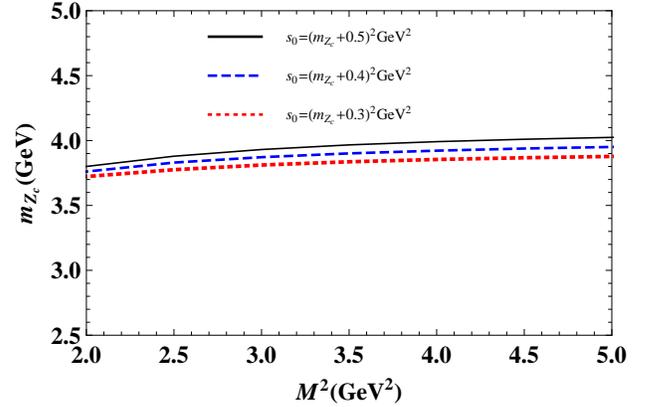}}
\end{picture}
}
\caption{The mass $m_{Z_{c}}$ as a function of the Borel parameter $M^2$ for
different values of $s_0$.}
\label{fig:Zmass}
\end{figure}

\begin{figure}
\centerline{
\begin{picture}(210,170)(0,0)
\put(-15,5){\epsfxsize8.2cm\epsffile{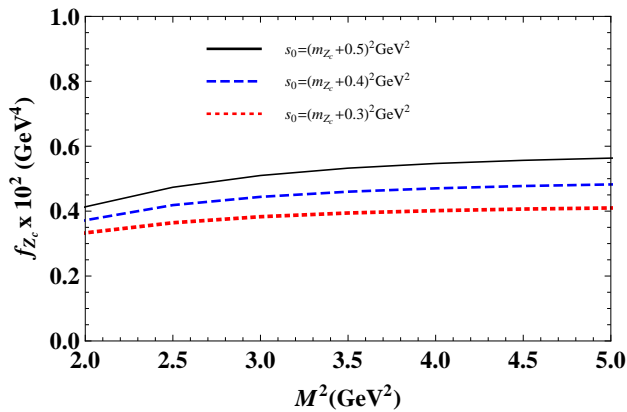}}
\end{picture}
}
\caption{The decay constant $f_{Z_{c}}$ vs Borel parameter $M^2$. The values
of the parameter $s_0$ are shown in the figure.}
\label{fig:Zcoupl}
\end{figure}

The QCD sum rules expressions for the mass and decay constant of the $Z_{c}$
state, as well as ones for the couplings $g_{Z_{c}J/\psi \pi }$ and $%
g_{Z_{c}\eta _{c}\rho }$ contain numerous parameters that should be fixed in
accordance with the standard prescriptions. Thus, for numerical computation
of the $Z_{c}$ state's mass and decay constant we need values of the quark
and gluon condensates. Apart from that QCD sum rules for the couplings
contain masses and decay constants of the heavy ($J/\psi $, $\eta _{c}$) and
light ($\pi $, $\rho $) mesons. The values of all these parameters are
collected in Table \ref{tab:Param}.

Sum rules calculations require fixing of the threshold parameter $s_{0}$ and
a region within of which it may be varied. For $s_{0}$ we employ
\begin{equation}
(3.9+0.3)^2\,\,\mathrm{GeV}^2 \leq s_{0}\leq(3.9+0.5)^2 \,\,\mathrm{GeV}^2.
\end{equation}
We also find the range $2\ \mathrm{GeV}^2<M^2<5\ \mathrm{GeV}^2$ as a
reliable one for varying the Borel parameter, where the effects of the
higher resonances and continuum states, and contributions of the higher
dimensional condensates meet all requirements of QCD sum rules calculations.
Additionally, in this interval the dependence of the mass and decay constant
on $M^2$ is stable, and we may expect that the sum rules give the correct
results. By varying the parameters $M^2$ and $s_{0}$ within the allowed
ranges, as well as taking into account ambiguities arising from other input
parameters we estimate uncertainties of the whole calculations. The results
for the mass $m_{Z_c}$ and decay constant $f_{Z_c}$ are drown as the
functions of the Borel parameter in Figs.~\ref{fig:Zmass} and \ref%
{fig:Zcoupl}, respectively. The sensitivity of the obtained predictions to
the choice of $s_0$ are also seen in these figures, where three different
values for $s_0$ are utilized. Our result for $m_{Z_c}$ together with the prediction of Ref.\ \cite{Wang:2013vex} are written down in
Table \ref{tab:DecW}. 
As it has been emphasized above, the exotic tetraquark $Z_{c}$ state was
observed by BESIII and Belle collaborations \cite%
{Ablikim:2013mio,Liu:2013dau}, which measured its mass and width. The experimental data on
the mass of $Z_c$ state are also shown in Table \ref{tab:DecW}. As is seen, experimental results for
$m_{Z_c}$ are rather precise ($\sim 0.15\, \%-0.20\, \% $), whereas theoretical prediction made in the present work suffers
from errors $\sim 5\,\% -10\, \%$, which are inherent to sum rules
calculations, and may be considered as acceptable in the case under
consideration. Our finding for  $m_{Z_c}$ is consistent with the
data, despite large theoretical uncertainties originating mainly from the
choice of the parameters $s_0$ and $M^2$.  It  also agrees with the theoretical prediction of Ref.\ \cite{Wang:2013vex}.

For the decay constant we get:
\begin{equation}
f_{Z_c}=(0.46\pm 0.03)\times 10^{-2}\,\, \mathrm{GeV}^{4}.
\end{equation}
The same quantity was also calculated in Ref.\ \cite%
{Wang:2013vex}, where the following prediction was made:
\begin{equation}
\lambda _{Z_{c}}=2.20_{-0.29}^{+0.36}\times 10^{-2}\,\,\mathrm{GeV}^{5}.
\end{equation}%
Taking into account differences in definitions of the decay constants, and
re-scaling $f_{Z_{c}}$ by the factor $m_{Z_{c}}$ in our case we get:
\begin{equation}
m_{Z_{c}}f_{Z_{c}}=(1.79\pm 0.12)\times 10^{-2}\,\,\mathrm{GeV}^{5}.
\end{equation}%
The discrepancy  between two result may be attributed to the more complicated
phenomenological part in QCD sum rules calculations employed in Ref.\ \cite%
{Wang:2013vex}.

The values for the mass and decay constant, as well as corresponding sum
rules formulas are used to evaluate the strong couplings $g_{Z_{c}J/\psi \pi
}$ and $g_{Z_{c} \eta_c \rho }$, and width of the decays $Z_{c}\rightarrow
J/\psi \pi $ and $Z_{c}\rightarrow \eta _{c}\rho $. In the evaluation of the
couplings we employ the same range for $s_{0}$ and $M^{2}$ as in
calculations of the mass and decay constant. For $g_{Z_{c}J/\psi \pi }$ we
find:
\begin{equation}
g_{Z_{c}J/\psi \pi }=(0.39\pm 0.06)\ \mathrm{GeV}^{-1}.
\end{equation}%
The width of the decay $Z_{c}\rightarrow J/\psi \pi$ can be obtained by
means of Eq.\ (\ref{eq:DW})
\begin{equation}
\Gamma (Z_{c} \rightarrow J/\psi \pi )=(41.9\pm 9.4)\ \mathrm{MeV}.
\end{equation}%

For the coupling $g_{Z_{c}\eta _{c}\rho }$ and width of the decay $%
Z_{c}\rightarrow \eta _{c}\rho $ we get%
\begin{equation}
g_{Z_{c}\eta _{c}\rho } =(1.39\pm 0.15)\ \mathrm{GeV}^{-1},  
\end{equation}
and
\begin{equation}
\Gamma (Z_{c} \rightarrow \eta _{c}\rho )=(23.8\pm 4.9)\ \mathrm{MeV}.
\label{eq:RhoDW}
\end{equation}

\begin{table}[tbp]
\renewcommand{\arraystretch}{1.5} \addtolength{\arraycolsep}{3pt}
\begin{tabular}{|c|c|}
\hline\hline
\mbox {} & $m_{Z_c}$\,(\mbox{MeV}) \, 
   \\ \hline \mbox{BESIII}\,\,\,
\cite{Ablikim:2013mio} & $3899 \pm 6$  \\ \hline
\mbox{Belle}\,\,\, \cite{Liu:2013dau} & $3895 \pm 8$  \\
\hline \mbox{Present Work} & $3900\pm 210$  
\\ \hline
\mbox{Z.~Wang, T.~Huang}\, \cite{Wang:2013vex} & $3910^{+110}_{-90}$  \\
\hline\hline
\end{tabular}%
\renewcommand{\arraystretch}{1} \addtolength{\arraycolsep}{-1.0pt}
\caption{Experimental data and theoretical predictions for the mass of $%
Z_{c}$ state.}
\label{tab:DecW}
\end{table}

The decay width of the channels $Z_{c}\rightarrow J/\psi \pi $ and $%
Z_{c}\rightarrow \eta _{c}\rho $ were also analyzed in Ref.\ \cite%
{Dias:2013xfa}. The authors considered the $Z_{c}$ state as the isospin $1$
partner of the $X(3872)$, and therefore within SU(2) symmetry the masses of
two particles were accepted being equal exactly to each other. In other
words, here independent computation of the mass $m_{Z_{c}}$ was not carried
out. The couplings $g_{Z_{c}J/\psi \pi }$ and $g_{Z_{c}\eta _{c}\rho }$ were
extracted from the three-point correlation function using standard methods
of QCD sum rules. They obtained the value 
$\Gamma (Z_{c}\rightarrow J/\psi \pi )=29.1\pm 8.2\,\,\mathrm{MeV}$ 
for the width in $J/\psi \pi$ channel, which  is lower than  our prediction.
 At the same time, the decay width
\begin{equation}
\Gamma (Z_{c}\rightarrow \eta _{c}\rho )=27.5\pm 8.5\,\,\mathrm{MeV},
\end{equation}%
is in accord with our result [see, Eq.\ (\ref{eq:RhoDW})].

Considering the transitions $Z_{c}\rightarrow J/\psi \pi $ and $%
Z_{c}\rightarrow \eta _{c}\rho $ as dominant channels we obtain for the total width of $Z_{c}$ approximately
\begin{equation}
\Gamma_{Z_{c}}=65.7\pm 10.6\,\,\mathrm{MeV},
\end{equation}%
which is in a good consistency with Belle data $\Gamma_{Z_{c}}=63\pm 35\,\,\mathrm{MeV}$ \cite{Liu:2013dau} and prediction of Ref.\ \cite%
{Dias:2013xfa}, i.e.  $\Gamma_{Z_{c}}=63.0\pm 18.1\,\,\mathrm{MeV}$. Our result is also compatible, within the errors, with 
the data  $\Gamma_{Z_{c}}=46\pm 22\,\,\mathrm{MeV}$ \cite{Ablikim:2013mio} extracted by BESIII collaboration. The ratio of width in $\eta _{c}\rho$ channel to that
 of  $J/\psi \pi$ is equal to $0.56\pm 0.24$ which is in agreement with  the prediction  of \cite{Esposito}, 
obtained using a type II tetraquark model, i.e. within the color-spin Hamiltonian approach by neglecting the spin-spin interaction outside the diquarks.

In conclusion, we have employed QCD two-point sum rule to calculate the
mass and decay constant of the exotic $Z_c$ state. The obtained results have
been used as input information for studying the strong vertices $%
Z_cJ/\psi\pi $ and $Z_c\eta_c\rho$, and evaluating of the couplings $%
g_{Z_cJ/\psi\pi}$ and $g_{Z_c\eta_c\rho}$. To this end, we have utilized the
methods of QCD sum rules on the light-cone and soft-meson approximation.
Finally, the widths of the decays channels $Z_c\to J/\psi\pi$ and $Z_c\to
\eta_c\rho$ have been found. The theoretical framework used here is rather
simple and allows one to calculate the relevant quantities in terms a few
local matrix elements of the pion and $\rho$ meson. Its accuracy was checked
 by explicit calculations in Ref.\ \cite{Braun:1995}, where  the
strong $D^{\ast}D\pi$ and $B^{\ast}B\pi$ vertices were explored in the
full LCSR approach, and its soft $q \to 0$ limit. It was demonstrated
that both computational schemes lead to almost identical predictions for the
strong couplings $g_{D^{\ast}D\pi}$ and $g_{B^{\ast}B\pi}$. The soft-meson(pion)
approximation was also successfully employed to investigate some other processes
\cite{Braun:2006td,Braun:2007pz}.

Our results for the mass and total  width  of $Z_c$  are
consistent with available experimental data as well as other theoretical predictions.  The ratio of width in $\eta _{c}\rho$ channel to that
 of  $J/\psi \pi$  is also in agreement with  the prediction  of \cite{Esposito}, 
obtained via a different approach. The observed discrepancy between our result for $f_{Z_c}$ and that of Ref.\ \cite%
{Wang:2013vex} can be connected  with accuracy and features of the
used approaches. Further experimental measurements and theoretical
computations on parameters of the exotic states may help us to improve schemes and methods for
their investigations.

\section*{ACKNOWLEDGEMENTS}

S.~S~A. is grateful to T.~M.~ Aliev and V.~M.~Braun for numerous useful
discussions. S.~S.~A. also thanks colleagues from the Physics Department of
Kocaeli University for warm hospitality. The work of S.~S.~A. was supported
by the TUBITAK grant 2221-"Fellowship Program For Visiting Scientists and
Scientists on Sabbatical Leave". This work was also supported in part by
TUBITAK under the grant no: 115F183.

\appendix*

\section{A}

\renewcommand{\theequation}{\Alph{section}.\arabic{equation}}

\label{sec:App} In this appendix we have collected the results of our
calculations of the spectral density
\begin{equation}
\rho ^{\mathrm{QCD}}(s)=\rho ^{\mathrm{pert}}(s)+\sum_{k=3}^{k=5}\rho
_{k}(s),  \label{eq:A1}
\end{equation}%
necessary for evaluation of the $Z_{c}$ meson mass $m_{Z_c}$ and its decay
constant $f_{Z_c}$ from the QCD sum rule. In Eq.\ (\ref{eq:A1}) by $\rho
_{k}(s)$ we denote the nonperturbative contributions to $\rho ^{\mathrm{QCD}%
}(s)$. Neglecting the terms proportional to light quark masses, the explicit expressions for $\rho ^{\mathrm{pert}}(s)$ and $\rho
_{k}(s)$ are presented below as the integrals over the Feynman parameters $z$
and $w$:
\begin{widetext}
\begin{eqnarray}
&&\rho ^{\mathrm{pert}}(s) =\frac{1}{384\pi ^{6}}\int_{0}^{1}dz%
\int_{0}^{1-z}dw\frac{r^{8}}{(w-1)}\left\{ wz^{2}\left[ swz(w+z-1)-m_{c}^{2}%
\frac{w+z}{r}\right] ^{2}\right.   \notag \\
&& \times \left. \left[ m_{c}^{2}\frac{w+z}{r}(4w(w-1)+3z(w-1)+3z^{2})
-swz(w+z-1)(3z^{2}+(w-1)(7w+3z))\right] \right\} \theta (L),
\end{eqnarray}%
\begin{equation}
\rho _{3}(s)=\frac{1}{16\pi ^{4}}\int_{0}^{1}dz%
\int_{0}^{1-z}dwm_{c}r^{5}wz(w+z-1)\left[ \langle u\overline{u}\rangle
w+\langle d\overline{d}\rangle z\right] \left[ m_{c}^{2}\frac{w+z}{r}%
-swz(w+z-1)\right] \theta (L),
\end{equation}%
\begin{eqnarray}
&&\rho _{4}(s) =-\frac{1}{36864\pi ^{4}}\langle \alpha _{s}\frac{GG}{\pi }%
\rangle \int_{0}^{1}dz\int_{0}^{1-z}dw\frac{wz^{2}r^{6}(w+z-1)^{2}}{(w-1)}%
\left\{ 90szw^{2}(w+z-1)^{2}\left[ 3z^{2}+(w-1)\right. \right.   \notag \\
&&\left. \left. \times (5w+3z)\right] +2m_{c}^{2}\frac{1}{r}\left\{
48z^{4}(z-1)^{5}+64w^{6}(z^{3}-1)+wz(z-1)\left[ -15+z(z-1)\right. \right.
\right.   \notag \\
&&\left. \left. \left. \times (-135+16z(z-1)^{2}(15z-4))\right] +w^{5}\left[
-61+16z(-7+z^{2}(19z-20))\right] +w^{4}\left[ 329+8z(-43+2z(3z-1)\right.
\right. \right.   \notag \\
&&\left. \left. \times (6-22z+13z^{2})\right) \right] +w^{2}(z-1)\left[
-15+339z+z^{2}(-405+32z(z-1)^{2}(17z-10))\right] +w^{3}\left[
-219+795z\right.   \notag \\
&&\left. \left. +2z^{2}(-249-344z+16z^{2}(63-66z+23z^{2}))\right] \right\}\theta (L)
\end{eqnarray}%
\begin{equation}
\rho _{5}(s)=\frac{m_{c}}{16\pi ^{4}}\int_{0}^{1}dz%
\int_{0}^{1-z}dwr^{5}w^{2}z^{2}(w+z-1)^{2}m_{0}^{2}\left[ \langle u\overline{%
u}\rangle w+\langle d\overline{d}\rangle z\right] \theta (L),
\end{equation}
\end{widetext}

\end{document}